\documentstyle[12pt,epsf]{article}
\def\hybrid{\topmargin 0pt      \oddsidemargin 0pt
	\headheight 0pt \headsep 0pt
	\textheight 9in         % US paper
	\textwidth 6.25in       % A4 paper
	\marginparwidth .875in
	\parskip 5pt plus 1pt   \jot = 1.5ex}

\catcode`\@=11
\def\marginnote#1{}
\newcount\hour
\newcount\minute
\newtoks\amorpm
\hour=\time\divide\hour by60
\minute=\time{\multiply\hour by60 \global\advance\minute by-\hour}
\edef\standardtime{{\ifnum\hour<12 \global\amorpm={am}%
	\else\global\amorpm={pm}\advance\hour by-12 \fi
	\ifnum\hour=0 \hour=12 \fi
	\number\hour:\ifnum\minute<10 0\fi\number\minute\the\amorpm}}
\edef\militarytime{\number\hour:\ifnum\minute<10 0\fi\number\minute}

\def\draftlabel#1{{\@bsphack\if@filesw {\let\thepage\relax
   \xdef\@gtempa{\write\@auxout{\string
      \newlabel{#1}{{\@currentlabel}{\thepage}}}}}\@gtempa
   \if@nobreak \ifvmode\nobreak\fi\fi\fi\@esphack}
	\gdef\@eqnlabel{#1}}
\def\@eqnlabel{}
\def\@vacuum{}
\def\draftmarginnote#1{\marginpar{\raggedright\scriptsize\tt#1}}

\def\draft{\oddsidemargin -.5truein
	\def\@oddfoot{\sl preliminary draft \hfil
	\rm\thepage\hfil\sl\today\quad\militarytime}
	\let\@evenfoot\@oddfoot \overfullrule 3pt
	\let\label=\draftlabel
	\let\marginnote=\draftmarginnote
   \def\@eqnnum{(\theequation)\rlap{\kern\marginparsep\tt\@eqnlabel}%
\global\let\@eqnlabel\@vacuum}  }

\def\numberbysection{\@addtoreset{equation}{section}
	\def\theequation{\thesection.\arabic{equation}}}

\def\underline#1{\relax\ifmmode\@@underline#1\else
	$\@@underline{\hbox{#1}}$\relax\fi}

\def\titlepage{\@restonecolfalse\if@twocolumn\@restonecoltrue\onecolumn
     \else \newpage \fi \thispagestyle{empty}\c@page\z@
	\def\thefootnote{\fnsymbol{footnote}} }

\def\endtitlepage{\if@restonecol\twocolumn \else  \fi
	\def\thefootnote{\arabic{footnote}}
	\setcounter{footnote}{0}}  %\c@footnote\z@ }
\catcode`@=12
\relax

\def\beq{\begin{equation}}
\def\eeq{\end{equation}}
\def\bea{\begin{eqnarray}}
\def\eea{\end{eqnarray}}
\def\nn{\nonumber}

\relax
\numberbysection
\hybrid
\begin{document}
\begin{titlepage}
\begin{center}
October~2001 \hfill    PAR--LPTHE 01/? \\[.5in]
{\large\bf Universality of 
 coupled Potts models}\\[.2in]
        {\bf Vladimir S.~Dotsenko (1), Jesper Lykke Jacobsen (2), \\
             Xuan Son Nguyen (1), and Raoul Santachiara (1)}\\[.2in]
        {\bf (1)} {\it LPTHE\/}\footnote{Unit\'e Mixte de Recherche CNRS
UMR 7589.},
        {\it  Universit{\'e} Pierre et Marie Curie, Paris VI\\
              Universit{\'e} Denis Diderot, Paris VII\\
              Bo\^{\i}te 126, Tour 16, 1$^{\it er}$ {\'e}tage \\
              4 place Jussieu,
              F-75252 Paris Cedex 05, France.}\\
        {\bf (2)} {\it Laboratoire de Physique Th\'eorique et Mod\`eles
                   Statistiques, \\
                   Universit\'e Paris-Sud,
                   B\^atiment 100, F-91405 Orsay, France.}
\end{center}

\vskip .15in
\centerline{\bf ABSTRACT}
\begin{quotation}

{\small We study systems of $M$ Potts models coupled by their local
energy density. Each model is taken to have a distinct number of
states, and the permutational symmetry $S_M$ present in the case
of identical coupled models is thus broken initially. The duality
transformations within the space of $2^M-1$ multi-energy couplings are
shown to have a particularly simple form. The selfdual manifold has
dimension $D_M = 2^{M-1}-1$. Specialising to the case $M=3$, we
identify a unique non-trivial critical point in the three-dimensional
selfdual space. We compare its critical exponents as computed from the
perturbative renormalisation group with numerical transfer matrix
results. Our main objective is to provide evidence that at the
critical point of three different coupled models the symmetry $S_3$ is
restored.}

\end{quotation}
\end{titlepage}

\newpage

\section{Introduction}
\label{sec:intro}

In the study of coupled models, and also of disordered models in their
replica formulation, the permutation group symmetry $S_M$ is supposed
to play an essential role \cite{ludwig,djlp,dns}. Namely, the interaction
part of the action for a set of $M$ identical coupled models
\beq
  A_{\rm int} \propto \int {\rm d}^2x \; g \sum_{a \neq b}
  \varepsilon_{a}(x)\varepsilon_{b}(x)
\eeq
is explicitly invariant with respect to any permutation of the
models. Here $a$ and $b$ are replica indices, and
$\{\varepsilon_a(x)\}$ designates the set of local energy operators.
In the lattice definition of such coupled models the interaction part
of the Hamiltonian takes a similar form, with only $\int {\rm d}^2x$
replaced by a summation and $\{\varepsilon_a(x)\}$ by an appropriate
lattice expression%
\footnote{In the case of the Potts model with nearest-neighbour
spins $\sigma^{(a)}_i$ and $\sigma^{(a)}_j$ in the replica $a$, one has
$\varepsilon_a(x) \sim 1-\delta \big(\sigma^{(a)}_i,\sigma^{(b)}_j \big)$.}.

When one introduces asymmetric couplings, by generalising the common
coupling constant $g$ to a matrix $g_{ab}$,
\beq
  A_{\rm int} \propto \int {\rm d}^2x \; \sum_{a \neq b}
  g_{ab} \varepsilon_{a}(x) \varepsilon_{b}(x)
  \label{A_int}
\eeq
a perturbative renormalisation group (RG) analysis reveals that the
$S_M$ symmetry is restored at the fixed point. A detailed study of
this scenario, within the $\epsilon$-type perturbative RG for
coupled Potts models, was carried out in \cite{ls}.
Supposing all components of $g_{ab}$ to stay of order $\epsilon$, their
initial values being all positive%
\footnote{In \cite{ls} the interaction part of the action
$A_{\rm int}$ (\ref{A_int}) was actually defined with an extra minus
sign, so that initially all the components of $g_{ab}$ were
taken to be negative.},
it was shown in \cite{ls} that the only non-trivial fixed point, having one
attractive direction, all other directions being repulsive, is that
with $g_{ab} \equiv g$.%
\footnote{Ref.~\cite{ls} also identified other fixed points, which
are related to the basic one (with all $g_{ab}$ equal and of the same
sign) by changing the sign of some of its components. This is equivalent
to switching the sign of a certain number of energy operators.
This last operation is a symmetry of the individual Potts models,
corresponding to their self-duality (note however such duality transformations
on the individual models are not directly related to the global duality
transformations on the entire coupled system to be discussed in
section~\ref{sec:dual}). This argument implies that the critical properties
of the various critical points classified in \cite{ls} should be equivalent.}
This type of restoration of the symmetry $S_M$ could be called
``soft universality''.

In this paper we are going to argue for a ``strong universality'' in the
criticality of coupled models. We shall mainly be interested in coupling 
$M=3$ different models%
\footnote{The duality transformations of section~\ref{sec:dual} are
however valid for general $M$.},
via (\ref{A_int}), with $\{\varepsilon_a\}$ belonging to Potts models
with different number of states $\{q_1,q_2,q_3\}$. This breaks the
permutational symmetry in a ``strong'' sense. Still, the RG
calculations show the existence of a single fixed point with all
$\{g_{ab}\}$ being positive, like it is the case for identical models.

To our knowledge, Simon \cite{simon} was the first to apply the RG
analysis to a set of different coupled Potts models. The most general
model studied by this author was that of $M_1$ Potts models with $q_1$
states and $M_2$ Potts models with $q_2$ states ($q_1 \neq q_2$), all
of them being coupled. After determining the fixed point structure, he
computed the dimensions of the spin operators, as well as the RG
equations for the energy operators to two-loop order. The effect of
disorder on these coupled systems was also analysed.

Here we generalise the RG calculations of \cite{simon} to the case of three
different coupled Potts models $(q_1 \neq q_2 \neq q_3)$. We shall compute
the dimensions of energy operators, with a special focus on the symmetry of
the theory, at the non-trivial fixed point that generalises the one found in
\cite{djlp} for three identical models. Within the space of couplings
$\{g_{ab}\}$, the new fixed point is stable in one direction and unstable in
the others, the topology of the RG flows being similar to those of
\cite{djlp}. But there is also one apparent difference: The permutational
symmetry has disappeared, the coupled models being different.

The purpose of this paper will be to provide evidence, by using various
methods, that at the fixed point of three different coupled models the
apparently lost symmetry $S_3$ is restored, implying a ``strong
universality''.

This symmetry restoration cannot be observed on the level of
the initial action (\ref{A_int}), as discussed above, nor is it visible in
the perturbative RG treatment, or in the Hamiltonian of the explicit lattice
realisation. This is because $\{\varepsilon_{a}\}$ are the energy operators of
different models, with different scaling dimensions in particular.

The way we shall check for the restoration of the symmetry is by looking at
the spectrum of scaling dimensions at the new fixed point, within the sector
of energy operators. Like in the case of identical models
\cite{ludwig,djlp,dns}, the RG analysis implies that the three energy
operators of the decoupled models
$\{\varepsilon_1(x),\varepsilon_2(x),\varepsilon_3(x)\}$ will rearrange as
three particular linear combinations so as to form the new primary operators
at the fixed point of the coupled models.

In the case of identical models the corresponding linear combinations are easy
to guess on symmetry grounds. The irreducible representations (irreps) of the
group $S_3$ in the basis
$\{\varepsilon_1(x),\varepsilon_2(x),\varepsilon_3(x)\}$ consist of a
(symmetric) singlet
\beq
  \varepsilon_{\rm S}(x)=\varepsilon_{1}(x)+\varepsilon_{2}(x)+
  \varepsilon_{3}(x)
  \label{symm}
\eeq
and an (antisymmetric) doublet
\beq
 \cases{
 \varepsilon_{\rm A_1}=\varepsilon_{1}(x)-\varepsilon_{2}(x) \cr
 \varepsilon_{\rm A_2}=\varepsilon_{1}(x)-\varepsilon_{3}(x) \cr}
 \label{antisym}
\eeq
that act as the new primary operators at the fixed point
\cite{ludwig,djlp,dns}. The fact that the operators $\varepsilon_{\rm A_1}$
and $\varepsilon_{\rm A_2}$ belong to the same two-dimensional irrep means
that their dimensions must coincide:
$\Delta(\varepsilon_{\rm A_1})=\Delta(\varepsilon_{\rm A_2})$.
These are however in general different from the dimension
$\Delta(\varepsilon_{\rm S})$ of the one-dimensional irrep.

When coupling different models, the corresponding linear
combinations of $\varepsilon_{1}, \varepsilon_{2}, \varepsilon_{3}$ will
have more complicated coefficients which have to be calculated by the
RG technique. One will find something of the form:
\beq
 \cases{\varepsilon^{\ast}_{1}=K_{11}\varepsilon_{1}+K_{12}\varepsilon_{2}+
 K_{13}\varepsilon_{3}\cr
 \varepsilon^{\ast}_{2}=K_{21}\varepsilon_{1}+K_{22}\varepsilon_{2}+
 K_{23}\varepsilon_{3}\cr
 \varepsilon^{\ast}_{3}=K_{31}\varepsilon_{1}+K_{32}\varepsilon_{2}+
 K_{33}\varepsilon_{3}\cr}
 \label{lin-comb}
\eeq
Since the initial dimensions
$\Delta(\varepsilon_{1}),\Delta(\varepsilon_{2}), \Delta(\varepsilon_{3})$
of the decoupled models differ, one may expect that the critical dimensions
(RG eigenvalues) of the newly formed primary operators
$\varepsilon_1^*,\varepsilon_2^*,\varepsilon_3^*$ might all be different.

Our argument is that, in the case of coupling different models, it is not
the linear combinations (\ref{lin-comb}) which have to be examined
to analyse the
symmetry, but rather the spectrum of their critical dimensions.
Permuting $\varepsilon_{1}, \varepsilon_{2}, \varepsilon_{3}$
in the combinations (\ref{lin-comb}) does not make much sense
because they are different.
To permute $\varepsilon_{1}^{\ast}, \varepsilon_{2}^{\ast},
\varepsilon_{3}^{\ast},$ one first has to know their properties,
their scaling dimensions, to decide if it makes sense or not.

The conclusion will be that one has to study the spectrum of
dimensions at the new fixed point. This provides a representation 
independent information, independent of the way one has defined
the theory initially, as by its action (Hamiltonian) in Eq.~(\ref{A_int}).

If the symmetry $S_{3}$ is restored 
then the dimensions
\beq
 \Delta(\varepsilon^{\ast}_{1}), \ \Delta(\varepsilon^{\ast}_{2}), \
 \Delta(\varepsilon^{\ast}_{3})
\eeq
should form a singlet and a doublet, as is the case when one couples
initially identical models.

The rest of the paper is organised as follows.

In section~\ref{sec:RG} we analyse the case $M=3$ by perturbative RG
calculations. As will be explained towards the end of that section, in
this case the RG results alone are not sufficient to decide whether
the $S_3$ symmetry is restored or not. We therefore propose to study
the coupled models through a particular lattice realisation, following
\cite{djlp}. To that end, in section~\ref{sec:dual}, we extend the
duality transformations derived in \cite{Jacobsen} for
symme\-tri\-cally coupled models to the asymmetric case. The resulting
relations are valid for any number $M$ of coupled models, and for the
most general asymmetric multi-energy couplings. In particular we
determine the selfdual solutions, thus simplifying dramatically the
subsequent numerical simulations.

After recalling briefly, in section~\ref{sec:alg}, the most efficient transfer
matrix algorithm constructed in \cite{djlp}, we show how it may be generalised
to the case of asymmetrically coupled models. We then turn to the numerical
analysis in section~\ref{sec:num}. We locate the non-trivial critical point on
the selfdual manifold for various systems of $M=3$ coupled models, and we
provide accurate values of the central charge and the energetic scaling
dimensions. We shall find strong evidence for a singlet/doublet spectrum at
the new fixed point. This we interpret as a signal of the restoration of the
$S_{3}$ symmetry.

Finally, section~\ref{sec:disc} is devoted to remarks and conclusions.

\section{Renormalization group analysis}
\label{sec:RG}

In the continuum limit, the three coupled Potts models can be represented
by the action
\bea
  A           &=& \sum^3_{a=1} A^{(a)}_{0}+A_{\rm int} \\
  A_{\rm int} &=& \int {\rm d}^2 x \,
  \sum^3_{a \neq b} g^{0}_{ab} \varepsilon_{a}(x) \varepsilon_{b}(x)
  \label{A_int2}
\eea
Here $\{A_0^{(a)}\}$ represents three decoupled models, or more precisely the
three decoupled conformal field theories for these models at their respective
critical points. The interaction term $A_{\rm int}$ makes them coupled. In
general, the initial couplings $\{g^{0}_{ab}\}$ are taken to be different.
Since $\{\varepsilon_{a}\}$ have different dimensions when the models are
different, even if we start from identical couplings $g^{0}_{ab}=g_{0}$, they
will become different in the course of renormalisation.

We shall parametrise the dimensions of the energy operators
$\{\varepsilon_{a}\}$ as in the papers \cite{djlp,ls,simon,dpp1}:
\beq
 (\Delta^{(a)}_{\varepsilon})_{\rm phys} = 1 - \frac{3}{2} \epsilon_a
\eeq
where the physical dimension $(\Delta^{(a)}_{\varepsilon})_{\rm phys}$
corresponds to twice the conformal dimension, as usual. The quantity
$\epsilon_{a}$, which measures the deviation from the Ising model, appears
also in the Coulomb gas parameter
\beq
 (\alpha^{(a)}_{+})^{2}=\frac{4}{3}-\epsilon_a
\eeq
For the Ising model one has $\epsilon=0$, whence $\alpha_+^2=\frac43$ and
$(\Delta_\varepsilon)_{\rm phys}=1$. The parameter $\alpha_+^2$ is useful in
analytic calculations because it is simpler to use than the central charge of
the corresponding conformal theory.

The details of the RG calculations are the same as in Refs.~\cite{simon,dpp1}.
In particular, all the necessary integrals have been
calculated in these papers. The generalisation to the case of three different
models is straightforward, and we shall therefore only give the final results.

It turns out to be convenient to redefine the expansion parameters
$\{\epsilon_{a}\}$ and the coupling constants $\{g_{ab}\}$ as follows:
\beq
  \frac{3}{2} \epsilon_{a} = \tilde{\epsilon}_{a} \qquad
  g_{ab} = \frac{1}{4\pi} \tilde{g}_{ab}
  \label{conventions}
\eeq
{}From now on we shall adopt this convention, omitting the tildes throughout
for simplicity of notation.

The $\beta$-functions for the interaction~(\ref{A_int2}) are found in the
form (to two-loop order):
\beq
 \cases{
 \frac{{\rm d}g_{12}}{{\rm d}\xi} \equiv \beta_{12}(\{g_{ab}\}) =
 \epsilon_{12}g_{12}-g_{13}g_{23}-\frac{1}{2}g_{12}(g^{2}_{13}+g^{2}_{23}) \cr
 \frac{{\rm d}g_{13}}{{\rm d}\xi} \equiv \beta_{13}(\{g_{ab}\}) =
 \epsilon_{13}g_{13}-g_{12}g_{32}-\frac{1}{2}g_{13}(g^{2}_{12}+g^{2}_{32}) \cr
 \frac{{\rm d}g_{23}}{{\rm d}\xi} \equiv \beta_{23}(\{g_{ab}\}) =
 \epsilon_{23}g_{23}-g_{21}g_{31}-\frac{1}{2}g_{23}(g^{2}_{12}+g^{2}_{31}) \cr}
 \label{beta}
\eeq
Here $\xi$ is the RG parameter, $g_{ab}=g_{ba}$ by symmetry
of the action (\ref{A_int2}), and we have
defined $\epsilon_{ab}\equiv\epsilon_{a}+\epsilon_{b}$.

The renormalisation of the energy operators, to second order, is found to
be given by the equations
\beq
\frac{d\varepsilon_{a}}{d\xi}=-(1-\epsilon_{a})\varepsilon_{a}-
\sum_{b\neq a}g_{ab}\varepsilon_{b}-\frac{1}{2}(\sum_{d\neq a}(g_{ad})^{2})
\varepsilon_{a}
\eeq
In matrix form this reads
\bea
 \frac{{\rm d}\varepsilon_{a}}{{\rm d}\xi} &=& 
 -\sum^3_{b=1} \Delta_{ab} \varepsilon_{b} \\
 \Delta_{ab} &=& (1-\epsilon_{a})\delta_{ab}-\gamma_{ab}(\{g_{cd}\}) \\
 \gamma_{ab} &=& \left(\begin{array}{ccc} -\frac{1}{2}(g^{2}_{12}+g^{2}_{13})&
 -g_{12} & -g_{13}\\ -g_{12} & -\frac{1}{2}(g^{2}_{21}+g^{2}_{23}) & -g_{23}\\
 -g_{13} & -g_{23} & -\frac{1}{2}(g^{2}_{31}+g^{2}_{32})\end{array}\right)
\eea
To define the new primary operators and their scaling dimensions at the new
fixed point we shall have to diagonalise the matrix $\Delta_{ab}$. It is
however more convenient to regroup the terms
$-\epsilon_{a}\delta_{ab}-\gamma_{ab}$ so that
\bea
 \Delta_{ab}  &=& \delta_{ab}-\Lambda_{ab} \label{Lambda} \\
 \Lambda_{ab} &=& \left(\begin{array}{ccc}
 \epsilon_{1} - \frac{1}{2}(g^{2}_{12} + g^{2}_{13}) & -g_{12} & -g_{13} \\
 -g_{12} & \epsilon_{2} - \frac{1}{2}(g^{2}_{21} + g^{2}_{23}) & -g_{23} \\
 -g_{13} & -g_{23} & \epsilon_{3}-\frac{1}{2}(g^{2}_{31} + g^{2}_{32})
 \end{array}\right)
\eea
and we need only diagonalise the matrix $\Lambda_{ab}$.

The non-trivial zeros of the $\beta$-functions (\ref{beta}) are found as
\beq
 \cases{
 g^{\ast}_{12} = \sqrt{\epsilon_{13}\epsilon_{23}}\{1-\frac{1}{2}\epsilon_{12}
 -\frac{1}{4}(\epsilon_{13}+\epsilon_{23})\} \cr
 g^{\ast}_{13} = \sqrt{\epsilon_{12}\epsilon_{23}}\{1-\frac{1}{2}\epsilon_{13}
 -\frac{1}{4}(\epsilon_{12}+\epsilon_{23})\} \cr
 g^{\ast}_{23} = \sqrt{\epsilon_{12}\epsilon_{13}}\{1-\frac{1}{2}\epsilon_{23}
 -\frac{1}{4}(\epsilon_{12}+\epsilon_{13})\} \cr}
 \label{zeros}
\eeq
They correspond to the non-trivial fixed point of the coupled models which
we are interested in. It is readily checked that the fixed point (\ref{zeros})
is stable in one direction and unstable in the two others, as is the case when
one couples identical models. Furthermore, the topology of the RG flows is
similar.

Substituting the fixed point values of the couplings in the matrix
$\Lambda_{ab}$ and diagonalising it, one obtains, after some algebra,
the following expressions for the eigenvalues:
\bea
\lambda_{1}   &=& -\frac{a+b+c}{2}+\frac{3abc}{a+b+c} \label{lambda1} \\
\lambda_{2,3} &=& \frac{a+b+c}{2} -
\frac{6abc+a^{2}b+ab^{2}+a^{2}c+ac^{2}+b^{2}c+bc^{2}}{2(a+b+c)} \nonumber \\
&\pm& \frac{1}{2(a+b+c)} \left\{ 6a^{2}b^{2}c^{2}-2abc(a^{2}b+ab^{2}+
a^{2}c+ac^{2}+b^{2}c+bc^{2}+a^{3}+b^{3}+c^{3}) \right. \nonumber\\
&+& \left. a^{2}b^{2}(a+b)^{2}+a^{2}c^{2}(a+c)^{2}+b^{2}c^{2}(b+c)^{2}
\right\}^{1/2}
\label{lambda23}
\eea
%\bea
% \lambda_{1}   &=& -\frac{a+b+c}{2}+\frac{3abc}{a+b+c} \\
% \lambda_{2,3} &=& \frac{a+b+c}{2} -
% \frac{(a+b)(a+c)(b+c) + 4abc}{2(a+b+c)} \\
% &\pm& \frac{\sqrt{(a+b)^2(a+c)^2(b+c)^2-4abc \left[a^3+b^3+c^3+
% 2(a+b)(a+c)(b+c)-3abc \right]}}{2(a+b+c)} \nonumber
%\eea
where we have simplified the notation by means of the abbreviations
$a=\epsilon_{12}$, $b=\epsilon_{13}$ and $c=\epsilon_{23}$.
According to the definition (\ref{Lambda}) of $\Lambda_{ab}$,
the dimensions of the new primary operators $\{\varepsilon^{\ast}_a\}$,
cf.~Eq.~(\ref{lin-comb}), are related to the above eigenvalues through
\beq
 \Delta(\varepsilon^{\ast}_a) = 1- \lambda_{a}
 \label{dims}
\eeq
for $a=1,2,3$.
We recall that all these calculations have been done to second order
in $\epsilon$.

It should be remarked that in the special case of two of the parameters
being equal, $b=c$ and $a \neq b$, the expressions for the dimensions
simplify considerably. One finds:
\bea
 \Delta(\varepsilon^{\ast}_{1}) &=& 1+
 \frac{a+2b}{2}-\frac{3ab^{2}}{a+2b} \\
 \Delta(\varepsilon^{\ast}_{2},\varepsilon^{\ast}_{3}) &=& 1-\frac{a+2b}{2}+
 \frac{b(4ab+a^{2}+b^{2})}{a+2b}\pm\frac{b^{2}(a-b)}{a+2b}
\eea

At first order in $\epsilon$, the RG results for the dimensions
\bea
 \Delta(\varepsilon^{\ast}_{1}) &=& 1+\frac{a+b+c}{2} \label{order1a} \\
 \Delta(\varepsilon^{\ast}_{2})=\Delta(\varepsilon^{\ast}_{3})
 &=& 1-\frac{a+b+c}{2}
 \label{order1b}
\eea
form a singlet and a doublet, in accordance with the scenario for the
restoration of the $S_{3}$ symmetry discussed in the Introduction.
The degeneracy of the doublet is however lifted at order $\epsilon^2$,
by the last term in (\ref{lambda23}). This is true even when two of
the models are identical, and only for $a=b=c$ does one recover the
degeneracy
$\Delta(\varepsilon^{\ast}_{2})=\Delta(\varepsilon^{\ast}_{3})$ \cite{djlp}.

In section~\ref{sec:num}, where the results of our numerical work are
presented, we shall show that this last result of the perturbative RG is wrong%
\footnote{We would still claim, of course, that our RG calculations are
technically correct!}.
We provide evidence that the splitting between the dimensions of
$\varepsilon^{\ast}_{2}$ and $\varepsilon^{\ast}_{3}$ is actually zero, and
that the symmetry restoration scenario therefore holds true.

We could suggest the following argument for the failure of the perturbative
RG. The $\epsilon$-expansion calculations are valid for perturbed conformal
theories because they respect the conformal symmetry, just as dimensional
regularisation is valid in the context of pertubative calculations in a gauge
theory because the method respects the gauge symmetry (otherwise the results
would be dependent on the regularisation technique). In the present problem,
the $\epsilon$-regularisation should be correct as far as the conformal
symmetry alone is concerned. But in case of extra symmetries, such as $S_{3}$,
the method might well give wrong results.

The RG formulae (\ref{lambda1})--(\ref{dims})
for the scaling dimensions, and for the central
charge which will be given below, are still quite useful. Apart from the
degeneracy issue just discussed, they compare well with the numerical results
of section~\ref{sec:num} when $\{\epsilon_a\}$ are small enough.

As far as the dimensions $\Delta(\varepsilon^{\ast}_{2}), \Delta(\varepsilon
^{\ast}_{3})$, defined by Eqs.~(\ref{lambda23})--(\ref{dims}),
are concerned, it is their mean value which compares
well with numerical results. It is worth noticing that the splitting,
namely the third term in (\ref{lambda23}), is numerically smaller than
the principal $\epsilon^{2}$ term, the second term in (\ref{lambda23}).
This is because of the compensation of negative and positive terms in the
third term of (\ref{lambda23}).%
\footnote{It is easy to check that the expression under the square-root
sign of (\ref{lambda23}) is always non-negative, so that $\lambda_{2,3}$
are well defined.}

The central charge can be obtained in a simple way using Zamolodchikov's
$c$-theorem \cite{ctheorem}. This theorem provides us with a function of the
couplings $c(g_{12},g_{13},g_{23})$ which decreases along the renormalisation
flow and takes a value $c(g_{12}^{*},g_{13}^{*},g_{23}^{*})$ at the fixed
point of the flow which equals the central charge of the associated conformal
field theory.
 
With the conventions (\ref{conventions}) taken into account, the $c$-function
is uniquely determined by
\bea
 \frac{\partial c(g_{12},g_{13},g_{23})}{\partial g_{ab}} &=&
 -\frac{3}{2} \beta_{ab}(g_{12},g_{13},g_{23}) \label{betadif} \label{ccc1} \\
 c(0,0,0) &=& c_0 \equiv c_1 + c_2 + c_3 \label{ccc2}
\eea
where $c_0$ is the total central charge of three decoupled models.

{}From Eq.~(\ref{beta}) and Eqs.~(\ref{ccc1})--(\ref{ccc2})
the $c$-function turns out to be given by:
\beq
  c(g_{12},g_{13},g_{23}) = c_{0} -
  \frac{3}{4}(a g_{12}^{2}+b g_{13}^{2}+c g_{23}^{2})
  +\frac{3}{2} g_{12}g_{13}g_{23}
  +\frac{3}{8}(g_{12}^{2}g_{13}^{2}+ g_{12}^{2}
  g_{23}^{2}+g_{13}^{2}g_{23}^{2})
  \label{c_func}
\eeq
At the fixed point we insert Eq.~(\ref{zeros}) into Eq.~(\ref{c_func});
up to the order $\epsilon^{4}$ the correction $\Delta c$ of the central
charge is:
\beq
 \Delta c =  -\frac{3}{4}abc +\frac{3}{8}(abc^{2}+bca^{2}+cab^{2})
 \label{c_RG}
\eeq

\section{Duality transformations}
\label{sec:dual}

The possibility of endowing the Potts model with a duality transformation
was one of the main motivations for introducing it \cite{Potts53}.
The study of duality for several Potts models coupled by their local
energy density was initiated by Domany and Riedel, who worked out the
case of two models with $q_1$ and $q_2$ states \cite{Domany}.
Technically, these authors used a particular version of the method of
lattice Fourier transforms \cite{Savit}, due to Wu and Wang \cite{Wu76}.
Dotsenko {\em et al.} generalised the computations to three symmetrically
coupled $q$-state models \cite{djlp}. However, it became clear that
the complexity of the Fourier method grew rapidly as the number of
models to be coupled was increased. Recently, it was pointed out by
Jacobsen that trading the Potts spin variables for a formulation in
terms of random clusters \cite{Kasteleyn}, or loops \cite{Baxter82},
the duality relations simplified dramatically. This observation made
it possible to work out the case of $M$ coupled $q$-state Potts models,
with the most general coupling by local energy densities consistent with
an $S_M$ symmetry \cite{Jacobsen}.

We now show how to treat the even more general case, where each
model does not necessarily have the same number of states, and the coupling
by local energy densities is the most general one.

\subsection{General case}

Consider a set ${\cal L}^M$ of $M$ identical planar lattices ${\cal L}$, which
we imagine to be stacked on top of one another. On each lattice site $i \in
{\cal L}$, and for each layer $\mu = 1,2,\ldots,M$, we define a Potts spin
$\sigma^{(\mu)}_i$ taking the values $\sigma^{(\mu)}_i =
1,2,3,\ldots,q_\mu$. The layers interact by means of the reduced
Hamiltonian
\beq
 {\cal H} = \sum_{\langle ij \rangle} {\cal H}_{ij},
\eeq
where $\langle ij \rangle$ denotes the set of lattice edges, and the
nearest-neighbour interaction is defined as
\beq
 {\cal H}_{ij} = - \sum_{\ell \subset {\cal E}} K_{\ell}
 \prod_{\mu \in \ell} \delta \left( \sigma^{(\mu)}_i,\sigma^{(\mu)}_j \right).
 \label{Hamil}
\eeq
By definition, the Kronecker delta function $\delta(x,y) = 1$ if $x=y$, and
zero otherwise. We have here defined ${\cal E}$ as the product set
$\prod_{\mu=1}^M \{ \emptyset,{\cal L}^{(\mu)} \}$, so any one of the $2^M$
subsets in ${\cal E}$ can be interpreted as a certain subset of layer indices.
One can also think of $\ell \subset {\cal E}$ as specifying the {\em state} of
a given edge $\langle ij \rangle$, meaning that in the interaction term
(\ref{Hamil}), $\ell$ determines which of the layers contribute to the product
$\prod_{\mu \in \ell}$ of the corresponding delta functions. The layers
specified by some $\ell \subset {\cal E}$ then interact by means of the
product of their local energy densities, through a coupling constant
$K_{\ell}$.

For later convenience, we shall represent a subset $\ell \subset {\cal E}$ as
a list of $M$ open ($\circ$) or filled ($\bullet$) circles, the $\mu$th circle
indicating respectively the absence (or presense) of the factor $\delta \big(
\sigma^{(\mu)}_i,\sigma^{(\mu)}_j \big)$. Furthermore, we sometimes use a
single open (resp.~filled) circle as an abbreviation of a list of $M$ open
(resp.~filled) circles.

When $M=1$, the model defined by Eq.~(\ref{Hamil}) reduces to the conventional
Potts model, whilst for $M=2$ it is identical to the Ashkin-Teller like model
considered in Ref.~\cite{Domany}. For $M=3$, it is the asymmetric version of
the model discussed in Ref.~\cite{djlp}, which had
$q \equiv q_1=q_2=q_3$, and whose couplings possessed an $S_3$ symmetry
upon permutation of the layers. Using the above symbolic notation,
%\beq
% {\cal E} = \{ \circ\circ\circ,
%               \bullet\circ\circ,\circ\bullet\circ,\circ\circ\bullet,
%               \circ\bullet\bullet,\bullet\circ\bullet,\bullet\bullet\circ,
%               \bullet\bullet\bullet \},
% \label{colours}
%\eeq
this symmetry can be expressed by the identities
\bea
 K_1 &\equiv& K_{\bullet\circ\circ} = K_{\circ\bullet\circ}
            = K_{\circ\circ\bullet} \nn \\
 K_2 &\equiv& K_{\circ\bullet\bullet} = K_{\bullet\circ\bullet}
            = K_{\bullet\bullet\circ} \\
 K_3 &\equiv& K_{\bullet\bullet\bullet} \nn.
\eea

By means of a generalised Kasteleyn-Fortuin transformation \cite{Kasteleyn}
the local Boltzmann weights can be recast as
\beq
 \exp(-{\cal H}_{ij}) = \prod_{\ell \subset {\cal E}}
   \left[1 + \left( {\rm e}^{K_\ell}-1 \right)
   \prod_{\mu \in \ell} \delta \left( \sigma^{(\mu)}_i,\sigma^{(\mu)}_j \right)
 \right],
 \label{Boltzmann}
\eeq
where we have simply used that $\exp(K \delta) = 1+[\exp(K)-1]\delta$,
if $\delta$ can only take the values 0 and 1.
Since furthermore $\delta^2 = \delta$, expanding the product over $\ell$
will lead to the equivalent form
\beq
 \exp(-{\cal H}_{ij}) = b_\circ + \sum_{\ell \subset {\cal E}} b_\ell
 \prod_{\mu \in \ell} \delta \left( \sigma^{(\mu)}_i,\sigma^{(\mu)}_j \right),
 \label{Boltzmann1}
\eeq
defining the coefficients $b_\ell$. The normalisation of Eq.~(\ref{Boltzmann})
is expressed by the fact that $b_\circ = 1$.

To relate the $b_\ell$ to the physical coupling constants $K_\ell$,
we evaluate Eqs.~(\ref{Boltzmann}) and (\ref{Boltzmann1}) in the situation
where $\delta(\sigma^{(\mu)}_i,\sigma^{(\mu)}_j)=1$ for $\mu \in \ell'$,
and zero otherwise. This resulting equations
\beq
 \exp\left(\sum_{\rho \subset \ell'} K_\rho \right)
 = \sum_{\rho \subset \ell'} b_\rho
\eeq
can readily be solved by applying the principle of inclusion-exclusion
\cite{Birkhoff}, yielding
\beq
 b_\ell = \sum_{\ell' \subset \ell} (-1)^{|\ell| - |\ell'|}
          \exp \left( \sum_{\rho \in \ell'} K_\rho \right).
 \label{b-from-K}
\eeq

The partition function in the spin representation
\beq
 Z = \sum_{\lbrace \sigma \rbrace}
     \prod_{\langle ij \rangle} \exp(-{\cal H}_{ij})
\eeq
can now be transformed into the random cluster representation as follows.
First, insert Eq.~(\ref{Boltzmann1}) on the right-hand side of the above
equation, and imagine expanding the product over the lattice edges $\langle ij
\rangle$. To each term in the resulting sum we associate an edge colouring
${\cal G}$ of ${\cal L}^M$, where an edge $\langle ij \rangle$ in layer $\mu$
is considered to be coloured (occupied) if the term contains the factor
$\delta(\sigma^{(\mu)}_i,\sigma^{(\mu)}_j)$, and uncoloured (empty) if it does
not. In other words, an edge colouring ${\cal G}$ is determined by specifying
an edge state $\ell_{ij} \subset {\cal E}$ for every edge $\langle ij \rangle
\in {\cal L}$.

The summation over the spin variables $\{ \sigma \}$ is now trivially
performed, yielding a factor of $q_\mu$ for each connected component
(cluster) in layer $\mu$ of the colouring graph. Keeping track of the
prefactors multiplying the $\delta$-functions, using Eq.~(\ref{Boltzmann1}),
we conclude that
\beq
 Z = \sum_{\cal G} \prod_{\ell \subset {\cal E}} b_\ell^{B_\ell}
     \prod_{\mu \in \ell} q_\mu^{C_\mu},
 \label{Z-cluster}
\eeq
where $C_\mu$ is the number of clusters in layer $\mu$, and $B_\ell$
is the number of edges $\langle ij \rangle \in {\cal L}^M$ having the
state $\ell$.

It is worth noticing that the random cluster description of the model has the
advantage that the $q_\mu$ only enter as parameters. By analytic
continuation one can thus give meaning to a non-integer number of states. The
price to be paid is that the $C_\mu$ are, a priori, non-local quantities.

In terms of the edge variables $b_\ell$ the duality transformation of the
partition function is easily worked out. For simplicity we shall assume that
the couplings constants $K_\ell$ are identical between all nearest-neighbour
pairs of spins. The generalisation to an arbitrary inhomogeneous distribution
of couplings is trivial; it suffices to let $K_\ell$ depend on $\langle ij
\rangle$ in Eq.~(\ref{Boltzmann}).

By definition, we take the colouring state $\ell \subset {\cal E}$ of the edge
$\langle ij \rangle \in {\cal L}$ to be dual to the complementary colouring
$\ell^*$ of its intersecting dual edge $\widetilde{\langle ij \rangle} \in
\tilde{\cal L}$. In the symbolic notation introduced above, the complementarity
operation $*$ simply means replacing every $\bullet$ by a $\circ$, and vice
versa. Also, note that we refer to dual quantities by a tilde throughout.

To establish the duality transformations, we begin by postulating that the
configuration ${\cal G}_{\bullet}$ with all lattice edges coloured, be dual to
the configuration ${\cal G}_{\bullet}^* \equiv {\cal G}_{\circ}$ with no
coloured (dual) edge.

This requirement fixes the constant entering the duality transformation.
Indeed, from Eq.~(\ref{Z-cluster}), we find that ${\cal G}_{\bullet}$ has
weight $b_{\bullet}^E \prod_{\mu=1}^M q_\mu$, where $E$ is the total
number of lattice edges, and ${\cal G}_{\circ}$ is weighted by
$\tilde{b}_{\circ}^E \prod_{\mu=1}^M q_\mu^F$, where $F$ is the number
of faces, including the exterior one.
We thus seek for a duality transformation of the form
\beq
 \frac{Z(\{ b_\ell \})}{\tilde{Z}(\{ \tilde{b}_\ell \})} =
 \left( \tilde{b}_{\circ} / b_{\bullet} \right)^E
 \prod_{\mu=1}^M q_\mu^{F-1},
\eeq
where for any configuration ${\cal G}$ the edge weights must transform so as
to keep the same {\em relative} weight between ${\cal G}$ and ${\cal
G}_{\bullet}$ as between ${\cal G}^*$ and ${\cal G}_{\circ}$.

An arbitrary colouring configuration ${\cal G}$ entering Eq.~(\ref{Z-cluster})
can be generated by applying a finite number of changes to ${\cal
G}_{\bullet}$, in which an edge of weight $b_\bullet$ is changed into an edge
of weight $b_\ell$ for some $\ell \subset {\cal E}$. By such a change, in
general, a pivotal bond is removed from the colouring graph in each of some
subset $\ell' \subset \ell^*$ of layers, thus creating $|\ell'|$ new clusters
in the corresponding layers. The weight relative to that of ${\cal
G}_{\bullet}$ will therefore change by $(b_\ell / b_\bullet) \prod_{\mu \in
\ell'} q_\mu$. On the other hand, in the dual configuration $\tilde{\cal G}$ a
cluster will be lost in each of the layers $(\ell')^* \cap \ell^*$, since each
of the $|\ell'|$ new clusters mentioned above will be accompanied by the
formation of a {\em loop} in $\tilde{\cal G}$. The weight change relative to
${\cal G}_{\circ}$ therefore amounts to $(\tilde{b}_{\ell^*}/\tilde{b}_0)
\prod_{\mu \in (\ell')^* \cap \ell^*} q_\mu^{-1}$. Comparing these two changes
we see that the factor $\prod_{\mu \in \ell'} q_\mu$ cancels nicely, and the
duality transformation takes the simple form
\beq
 \tilde{b}_\ell = (b_{\ell^*}/b_\bullet) \prod_{\mu \in \ell} q_\mu,
 \label{dual}
\eeq
the relation with $\ell = \circ$ being trivial.

Note in particular that Eq.~(\ref{dual}) with $\ell = \bullet$ implies that
$b_\bullet \tilde{b}_\bullet = \prod_{\mu=1}^M q_\mu$, since $b_\circ=1$
by definition. Then, dualising Eq.~(\ref{dual}) once again yields
\beq
 \widetilde{\widetilde{b_\ell}} =
 \frac{1}{\tilde{b}_\bullet} \prod_{\mu \in \ell} q_\mu \cdot
 \frac{b_\ell}{b_\bullet} \prod_{\mu \in \ell^*} q_\mu = b_\ell,
\eeq
so that the duality transformation is indeed involutive, as required.

The duality relations (\ref{dual}) can be recast in an even simpler form
by trading the random clusters for the loops surrounding them (and their
duals) on the lattice ${\cal L}_{\rm m}$ medial to ${\cal L}$ \cite{Baxter82}.
Using the Euler relation, we find that Eq.~(\ref{Z-cluster}) must be
replaced by
\beq
 Z = \left( \prod_{\mu=1}^M q_\mu \right)^{N/2}
     \sum_{\cal G} \prod_{\ell \subset {\cal E}} x_\ell^{B_\ell}
     \prod_{\mu \in \ell} q_\mu^{L_\mu/2},
 \label{Z-loop} 
\eeq
where, by a slight abuse of notation, we use the same notation ${\cal G}$ for
the loop and the cluster configurations, since they are in bijective
correspondence. $L_\mu$ are now the number of closed loops in layer $\mu$,
and $N$ is the total number of vertices in ${\cal L}$. Note that the
bond weights $b_\ell$ have now been replaced by
\beq
 x_\ell = b_\ell \prod_{\mu \in \ell} q_\mu^{-1/2}.
\eeq
It is easily verified that the duality relations (\ref{dual}) now simply
read
\beq
 \tilde{x}_\ell = x_{\ell^*}.
 \label{dual-loop}
\eeq
This is our main result.

In the case of a lattice ${\cal L} = \tilde{\cal L}$ which is unchanged by the
duality, such as the infinite square lattice, we can now search for selfdual
solutions. These are obtained by imposing $\tilde{b}_\ell = b_\ell$, and read
explicitly
\beq
 x_\ell^{\rm(s.d.)} = x_{\ell^*}^{\rm(s.d.)}.
 \label{selfdual-loop}
\eeq
Since $x_\circ = 1$ by normalisation, the selfdual manifold has dimension
$D_M = 2^{M-1}-1$.

Two special points always belong to the selfdual manifold.
The first one is
\beq
 x_\ell = 1, \ \ \ \ \mbox{\rm for any $\ell \subset {\cal E}$}.
 \label{special1}
\eeq
It is straightforward to verify that in terms of the original couplings this
means $K_{\circ \cdots \circ \bullet \circ \cdots \circ} = 1 + q_\mu^{1/2}$
(with the unique $\bullet$ at position $\mu$) for $\mu=1,2,\ldots,M$, all
other $K_\ell$ being zero. In other words, this is just $M$ non-interacting
selfdual (critical) Potts models.

The other special point is
\beq
 x_\circ = x_\bullet = 1, \ \ \ \ \mbox{\rm (all other $x_\ell = 0$)}.
 \label{special2}
\eeq
In terms of the original couplings this means $K_\bullet = 1 + \prod_{\mu=1}^M
q_\mu^{1/2}$, all other $K_\ell$ being zero. In this case, the $M$ models
effectively couple so as to form a {\em single} critical Potts model with
$\prod_{\mu=1}^M q_\mu$ states.

\subsection{Two models}

Let us briefly show how to recover the result for $M=2$ \cite{Domany}
from the compact formulation of Eq.~(\ref{dual-loop}).

Introducing the shorthand notation
$\delta_\mu = \delta \big( \sigma_i^{(\mu)},\sigma_j^{(\mu)} \big)$
for $\mu=1,2$,
the Hamiltonian (\ref{Hamil}) reads in this case
\beq
 -{\cal H}_{ij} =
   K_{\bullet \circ} \delta_1 + K_{\circ \bullet} \delta_2 +
   K_{\bullet \bullet} \delta_1 \delta_2.
\eeq
{}From Eq.~(\ref{b-from-K}) we have
\bea
 \sqrt{q_1} x_{\bullet \circ} \equiv b_{\bullet \circ}
   &=& {\rm e}^{K_{\bullet \circ}} - 1 \nn \\
 \sqrt{q_2} x_{\circ \bullet} \equiv b_{\circ \bullet}
   &=& {\rm e}^{K_{\circ \bullet}} - 1     \\
 \sqrt{q_1 q_2} x_{\bullet\bullet} \equiv b_{\bullet\bullet}
   &=& {\rm e}^{K_{\bullet\bullet}+K_{\bullet \circ}+K_{\circ \bullet}} -
  \left( {\rm e}^{K_{\bullet \circ}} + {\rm e}^{K_{\circ \bullet}} \right) + 1,
\eea
and using the selfduality criteria $x_{\bullet \circ} = x_{\circ \bullet}$ and
$x_{\bullet \bullet} = x_{\circ \circ} \equiv 1$ we readily find the solutions
\bea
 {\rm e}^{K_{\circ \bullet}}  &=&
   1 + \sqrt{\frac{q_2}{q_1}}
   \left( {\rm e}^{K_{\bullet \circ}}-1 \right) \nn \\
 {\rm e}^{K_{\bullet\bullet}} &=&
 \frac{\sqrt{q_1}+\sqrt{q_2} \left[ 1+(q_1-1){\rm e}^{-K_{\bullet \circ}}
 \right]}{\sqrt{q_1}+\sqrt{q_2} \left( {\rm e}^{K_{\bullet \circ}}-1 \right)}.
\eea

\subsection{Three models}

For the case $M=3$, the Hamiltonian (\ref{Hamil}) reads
\bea
 -{\cal H}_{ij} &=&
   K_{\bullet \circ \circ} \delta_1 + K_{\circ \bullet \circ} \delta_2 +
   K_{\circ \circ \bullet} \delta_3 + \nn \\
 & &  K_{\circ \bullet \bullet} \delta_2 \delta_3 +
   K_{\bullet \circ \bullet} \delta_1 \delta_3 +
   K_{\bullet \bullet \circ} \delta_1 \delta_2 +
   K_{\bullet \bullet \bullet} \delta_1 \delta_2 \delta_3.
\eea
{}From Eq.~(\ref{b-from-K}) we have
\bea
 b_{\bullet \circ \circ}
   &=& {\rm e}^{K_{\bullet \circ \circ}} - 1 \nn \\
 b_{\circ \bullet \bullet}
   &=& {\rm e}^{K_{\circ \bullet \bullet} + K_{\circ \bullet \circ} +
   K_{\circ \circ \bullet}} - 
   \left( {\rm e}^{K_{\circ \bullet \circ}} + {\rm e}^{K_{\circ \circ \bullet}}
   \right) + 1 \nn \\
 b_{\bullet\bullet\bullet}
   &=& {\rm e}^{K_{\bullet\bullet\bullet}+K_{\circ \bullet \bullet}+
   K_{\bullet \circ \bullet} + K_{\bullet \bullet \circ} +
   K_{\bullet \circ \circ} + K_{\circ \bullet \circ} + K_{\circ \circ \bullet}}
   - \nn \\
   & &\left(
   {\rm e}^{K_{\circ \bullet \bullet} + K_{\circ \bullet \circ} +
   K_{\circ \circ \bullet}} +
   {\rm e}^{K_{\bullet \circ \bullet} + K_{\bullet \circ \circ} +
   K_{\circ \circ \bullet}} +
   {\rm e}^{K_{\bullet \bullet \circ} + K_{\bullet \circ \circ} +
   K_{\circ \bullet \circ}} \right) + \nn \\
   & & \left(
   {\rm e}^{K_{\bullet \circ \circ}} + {\rm e}^{K_{\circ \bullet \circ}} +
   {\rm e}^{K_{\circ \circ \bullet}} \right) - 1,
\eea
where the first and second line each represent three equations that can be
obtained by cyclically rotating the layer indices.

Imposing the selfduality criteria
$\sqrt{q_1} \, b_{\circ \bullet \bullet}=
 \sqrt{q_2 q_3} \, b_{\bullet \circ \circ}$
and
$b_{\bullet \bullet \bullet} = \sqrt{q_1 q_2 q_3}$
we obtain after a little algebra
\bea
 {\rm e}^{K_{\circ \bullet \bullet}} &=& \frac{
 \sqrt{\frac{q_2 q_3}{q_1}} \left( {\rm e}^{K_{\bullet \circ \circ}}-1\right) +
 {\rm e}^{K_{\circ \bullet \circ}} + {\rm e}^{K_{\circ \circ \bullet}} - 1}
 {{\rm e}^{K_{\circ \bullet \circ} + K_{\circ \circ \bullet}}} \nn \\
 {\rm e}^{K_{\bullet \bullet \bullet}} &=&  \frac{A}{B} \cdot
 {\rm e}^{K_{\bullet\circ\circ}+K_{\circ\bullet\circ}+K_{\circ\circ\bullet}}
 \nn \\
 A &=& \left( {\rm e}^{K_{\bullet \circ \circ}} - 1 \right) q_2 q_3 +
       \left( {\rm e}^{K_{\circ \bullet \circ}} - 1 \right) q_1 q_3 +
       \left( {\rm e}^{K_{\circ \circ \bullet}} - 1 \right) q_1 q_2 + \nn \\
   & & \left( {\rm e}^{K_{\bullet\circ\circ}}+{\rm e}^{K_{\circ\bullet\circ}}+
        {\rm e}^{K_{\circ\circ\bullet}}-2 \right) \sqrt{q_1 q_2 q_3} +
       q_1 q_2 q_3 \nn \\
 B &=& \left[ \left(
       {\rm e}^{K_{\circ\bullet\circ}}+{\rm e}^{K_{\circ\circ\bullet}}
       -1 \right) \sqrt{q_1} +
       \left( {\rm e}^{K_{\bullet\circ\circ}} - 1 \right) \sqrt{q_2 q_3}
       \right] \times \nn \\
   & & \left[ \left(
       {\rm e}^{K_{\bullet\circ\circ}}+{\rm e}^{K_{\circ\circ\bullet}}
       -1 \right) \sqrt{q_2} +
       \left( {\rm e}^{K_{\circ\bullet\circ}} - 1 \right) \sqrt{q_1 q_3}
       \right] \times \nn \\
   & & \left[ \left(
       {\rm e}^{K_{\bullet\circ\circ}}+{\rm e}^{K_{\circ\bullet\circ}}
       -1 \right) \sqrt{q_3} +
       \left( {\rm e}^{K_{\circ\circ\bullet}} - 1 \right) \sqrt{q_1 q_2}
       \right].
\eea
These expressions generalise those given in Ref.~\cite{djlp}.

It should be clear that for higher $M$, manipulating such expressions
by direct use of the Fourier method \cite{Wu76} becomes extremely
cumbersome.

\section{Transfer matrix algorithm}
\label{sec:alg}

In Ref.~\cite{djlp} it was shown that the transfer matrix $T(L)$ for coupled
Potts models on semi-infinite strips of width $L$ may be written in a
variety of ways. For a given $L$, the smaller the dimension of $T(L)$ the
more efficient will be the computations, since both time and memory
consumption increase roughly linearly with ${\rm dim} \; T(L)$.

The best choice turns out to be to write the transfer matrix for the loop
model on the medial lattice ${\cal L}_{\rm m}$ \cite{Baxter82}, which was
referred to as algorithm ${\tt alg4}$ in \cite{djlp}. This also gives us the
advantage of having simple duality relations, cf.~Eq.~(\ref{dual-loop}), and
to be able to treat the numbers of states $\{q_1,q_2,q_3\}$ as continuous
parameters.

Let us recall from Eqs.~(\ref{Z-loop}) and (\ref{selfdual-loop}) that the
partition function for $M=3$ coupled models on the selfdual manifold can be
written, up to a trivial multiplicative constant, as
\beq
 Z = \sum_{\cal G}
 x_1^{B_{\bullet\circ\circ}+B_{\circ\bullet\bullet}}
 x_2^{B_{\circ\bullet\circ}+B_{\bullet\circ\bullet}}
 x_3^{B_{\circ\circ\bullet}+B_{\bullet\bullet\circ}}
 q_1^{L_1/2} q_2^{L_2/2} q_3^{L_3/2}
 \label{Z3}
\eeq
where we have put $x_1=x_{\bullet\circ\circ}$,
$x_2=x_{\circ\bullet\circ}$ and $x_3=x_{\circ\circ\bullet}$ for brevity.
For simplicity, we shall take the lattice ${\cal L}$ to be the square one,
so that ${\cal L}_{\rm m}$ is once again a square lattice.

\begin{figure}
\begin{center}
 \leavevmode
 \epsfysize=60pt{\epsffile{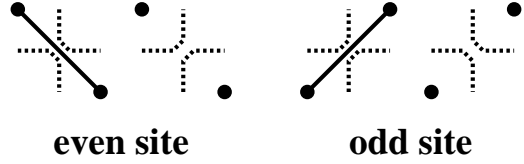}}
 \end{center}
 \protect\caption[3]{\label{fig:medial}The relation between the random
 cluster model on 
 the square lattice and the loop model on the corresponding medial
 graph. The clusters consist of connected components of coloured edges
 (thick lines) or isolated sites (filled circles). Loops on the medial
 graph (dashed lines) are defined by the convention that they wrap
 around the cluster boundaries.}
\end{figure}

In the loop picture, the symbols $\bullet$ and $\circ$ refer to the two
different ways of splitting the vertex, with a definition that alternates
between the even and the odd sublattice; see Fig.~\ref{fig:medial}. But thanks
to the selfduality, Eq.~(\ref{Z3}) is invariant with respect to a global
colour conjugation $\bullet \leftrightarrow \circ$, and so one might as well
forget about the distinction between the sublattices%
\footnote{Using the language of Ref.~\cite{Baxter82}, coupled Potts models
constitute a staggered vertex model, which however becomes homogeneous
exactly on the selfdual manifold.}.
The number of occurrences of a given vertex splitting (and hence the $B$'s)
can be counted locally, and are thus easily realised as local Boltzmann factors
in the transfer matrix.

The $L_\mu$ in (\ref{Z3}) count the number of closed loops in each layer
$\mu=1,2,3$. Despite of the non-local nature of the loops, these quantities do
not obviate the construction of the transfer matrix. Rather, they can be
counted locally by writing the transfer matrix in the basis of Catalan-like
connectivities, as described in \cite{djlp}. Formally, if ${\cal C}_k$
designates the space of pairwise Catalan connectivities among a set of
$L=2k$ points, $T(L)$ acts on the product space
${\cal S}_k = {\cal C}_k \bigotimes {\cal C}_k \bigotimes {\cal C}_k$.
One has ${\rm dim}\; {\cal C}_k = \frac{(2k)!}{k!(k+1)!}$.

Weighing the loops with a Boltzmann factor $\sqrt{q_\mu}$ that depends on
the layer $\mu$ is a trivial modification of \cite{djlp}. An important
difference, however, is that the layers are now distinguishable.
Accordingly, one has simply
${\rm dim} \; {\cal S}_k = ({\rm dim}\; {\cal C}_k)^3$.

As in Ref.~\cite{djlp} we transfer along one of the main directions of
the square lattice ${\cal L}_{\rm m}$, with periodic boundary conditions
in the transverse direction. To ensure that ${\cal C}_k$ be well defined,
we must take the strip width to be even, $L=2k$.

We shall also need to consider the situation where all vertex weights
$x$ tend to infinity in fixed ratios. Defining $x_2' = \frac{x_2}{x_1}$
and $x_3' = \frac{x_3}{x_1}$ as the relevant ratios, (\ref{Z3}) can
be rewritten, once again up to an irrelevant multiplicative constant,
as
\beq
 Z' = \sum_{\cal G'}
 x_2'^{B_{\circ\bullet\circ}+B_{\bullet\circ\bullet}}
 x_3'^{B_{\circ\circ\bullet}+B_{\bullet\bullet\circ}}
 q_1^{L_1/2} q_2^{L_2/2} q_3^{L_3/2}
 \label{Z3'}
\eeq
where the symbol ${\cal G'}$ means all colouring figurations in which
the local colourings $\circ\circ\circ$ and $\bullet\bullet\bullet$ do
not occur. In Ref.~\cite{djlp}, the special case $(x_2',x_3')=(1,1)$
with $q_1=q_2=q_3$ was identified as the non-trivial critical fixed
point for three identical coupled models.

\section{Numerical results}
\label{sec:num}

Using sparse matrix factorisation techniques, we have been able to numerically
compute the first few eigenvalues of the transfer matrices $T(L)$ for $Z$
(\ref{Z3}) and $T'(L)$ for $Z'$ (\ref{Z3'}) for even strip widths up to
$L_{\rm max}=12$. The largest matrices had dimension
${\rm dim} \; T(L_{\rm max}) = (132)^3$, but the sparse matrix factorisation
necessitates the use of intermediate states with $L+2$ dangling loop segments,
and so the largest sparse matrices involved were of dimension $(429)^3$.

\subsection{Phase diagram and central charge}

As a first test of our algorithm, we checked that it gave the correct
eigenvalues at the special points $(x_1,x_2,x_3)=(1,1,1)$ and $(0,0,0)$,
cf.~Eqs.~(\ref{special1}) and (\ref{special2}).

We then extracted the (effective) central charge from the leading eigenvalue
$\lambda_0$ in the standard way \cite{bcn,affleck}:
\beq
  f_0(L) = f_0(\infty) - \frac{\pi c}{6 L^2} + \cdots
  \label{cc}
\eeq
with $f_0(L) = - \frac{1}{L} \log \lambda_0$.
Using the $c$-theorem \cite{ctheorem}, we could readily establish the
topology of the RG flows:
\begin{itemize}
 \item In the space $(x_1,x_2,x_3)$, there is a one-parameter curve along
   which the partial derivatives of the effective central charge
   with respect to the two perpendicular directions vanish identically.
   Moreover, the corresponding second derivatives are strictly negative,
   so that this curve acts as a ``mountain ridge'' for $c_{\rm eff}$.
 \item The curve passes through the points $(0,0,0)$ and $(1,1,1)$, and
   then goes to infinity with some fixed ratios
   $(x_2',x_3') \equiv (x_2'^*,x_3'^*)$ that depend on $(q_1,q_2,q_3)$.
 \item For various values $(q_1,q_2,q_3) \in [2,4]^3$, with at most one
   $q_a = 2$,
   we observe that $c_{\rm eff}$ is a monotonically decreasing function when
   going along the curve from the point $(1,1,1)$ towards either
   $(0,0,0)$ or $x_1(1,x_2'^*,x_3'^*)$, $x_1 \to \infty$. In the former
   case, the decrease is rapid and gets more pronounced with increasing
   system size $L$, signalling the first-order nature ($c_{\rm eff}=0$)
   of the phase transition in the $(q_1 q_2 q_3)$-state Potts model.
   In the latter case, the decrease is very slight, distinguishing
   the point $(x_2'^*,x_3'^*)$ as a candidate for the non-trivial fixed
   point of three coupled Potts models in the lattice realisation.
\end{itemize}

Based on this evidence, we switched to the matrix $T'(L)$ in order to
accurately locate the point $(x_2'^*,x_3'^*)$ and study its critical
properties. Invoking again the $c$-theorem \cite{ctheorem}, this was
done by scanning the space $(x_2',x_3')$ for a maximum in $c_{\rm eff}$
for different system sizes. As usually \cite{djlp}, we obtained the
best precision by including a non-universal $1/L^4$ term in
(\ref{cc}), and so our fits are based on three consecutive strip widths $L$.
The obtained maximum can be interpreted as a finite-size pseudo-critical
point, which tends to $(x_2'^*,x_3'^*)$ as $L \to \infty$.

We have concentrated the main part of our computation time on two
important special cases: A tricritical Ising model and a three-state
Potts model coupled with either an Ising model or with a tricritical
three-state Potts model. In the language of minimal models, we refer
to these two situations as
${\cal M}_{345} \equiv {\cal M}_3 \times {\cal M}_4 \times {\cal M}_5$ and
${\cal M}_{456} \equiv {\cal M}_4 \times {\cal M}_5 \times {\cal M}_6$
respectively.

\begin{table}
\begin{center}
\begin{tabular}{l|llll}
 $L$     & $x_2'^*$   & $x_3'^*$   & $(c_{\rm eff})_{\rm max}$ 
         & $\Delta c$ \\ \hline
 2,4,6   & 0.8523(1)  & 0.7133(1)  & 1.93540138 & -0.00058   \\
 4,6,8   & 0.8476(1)  & 0.7136(1)  & 1.98040147 & -0.00352   \\
 6,8,10  & 0.8481(1)  & 0.7076(1)  & 1.99047341 & -0.00416   \\
 8,10,12 & 0.8483(1)  & 0.7061(1)  & 1.99348414 & -0.00443   \\
\end{tabular}
\end{center}
\caption{\label{tab:c345}Maxima of the effective central charge for the
 system ${\cal M}_{345}$, and the deviation from the value at the
 decoupling point.}
\end{table}

\begin{table}
\begin{center}
\begin{tabular}{l|llll}
 $L$     & $x_2'^*$   & $x_3'^*$   & $(c_{\rm eff})_{\rm max}$
         & $\Delta c$ \\ \hline
 2,4,6   & 0.9230(1)  & 0.8679(1)  & 2.27445933 & -0.0115    \\
 4,6,8   & 0.9218(1)  & 0.8661(1)  & 2.32198052 & -0.0185    \\
 6,8,10  & 0.9213(1)  & 0.8645(1)  & 2.33129236 & -0.0190    \\
 8,10,12 & 0.9213(5)  & 0.8645(5)  & 2.333272   & -0.0225    \\
% TO DO: Update L=12 result
\end{tabular}
\end{center}
\caption{\label{tab:c456}Maxima of the effective central charge for the
 system ${\cal M}_{456}$, and the deviation from the value at the
 decoupling point.}
\end{table}

The corresponding finite-size estimates for $(x_2'^*,x_3'^*)$
and for $(c_{\rm eff})_{\rm max}$ are given in Tables~\ref{tab:c345}
and \ref{tab:c456}. Rather than directly extrapolating
$(c_{\rm eff})_{\rm max}$ to the $L\to\infty$ limit, we consider instead
for any $L$ its deviation $\Delta c(L)$ with respect to the corresponding
value found at the point where the three Potts models decouple.
Since this deviation is numerically small, it is reasonable to expect that
the finite-size corrections to $c$ at $(x_2'^*,x_3'^*)$ are similar to those at
the decoupling point. Tables~\ref{tab:c345} and \ref{tab:c456} confirm
that this is indeed the case: the estimates for $\Delta c$ only depend
very weakly on $L$. As $L\to\infty$, we obtain finally the following
extrapolated values for $\Delta c$ and for the critical couplings:
\beq
 \begin{array}{lll}
 \mbox{Model ${\cal M}_{345}$}: &
 (x_2'^*,x_3'^*) = \big(0.8485(1),0.7053(5)\big) &
 \Delta c = -0.0050(3) \\
 \mbox{Model ${\cal M}_{456}$}: &
 (x_2'^*,x_3'^*) = \big(0.9210(5),0.862(2)\big) &
 \Delta c = -0.025(3) \\
 \end{array}
\eeq
% TO DO: Update L=12 for M_456.
The values of $\Delta c$
are seen to compare quite favourably with the two-loop result
of the perturbative RG (\ref{c_RG}), whose numerical values
read respectively
\beq
 \Delta c({\cal M}_{345})_{\rm RG} = -0.0041 \qquad
 \Delta c({\cal M}_{456})_{\rm RG} = -0.0175
\eeq
Actually, assuming that the RG series is alternating, the fifth-order
term not present in (\ref{c_RG}) would supposedly lead to even better
agreement with the numerical results.

\subsection{Higher exponents in the even sector}

Examining the higher eigenvalues of the transfer matrices, one has also
access to the scaling dimensions by using the formula \cite{Cardy83}
\beq
 f_i(L) - f_0(L) = \frac{2\pi (\Delta_i)_{\rm phys}}{L^2} + \cdots
 \label{scal_dims}
\eeq
We here work in the even sector of the transfer matrix, and besides
the identity operator (the free energy) we expect to find the various
energy-like primaries. In particular  we are interested in the primary
operators
$\varepsilon_{1}^{*}$, $\varepsilon_{2}^{*}$, $\varepsilon_{3}^{*}$
discussed in section~\ref{sec:RG}; cf.~Eq.~(\ref{lin-comb}).
According to Eqs.~(\ref{lambda1})--(\ref{dims}) we have  
$0<\Delta(\varepsilon_{2}^{*}),\Delta(\varepsilon_{3}^{*})< 1$ and 
$1<\Delta(\varepsilon_{1}^{*})<2$ for small enough $\{\epsilon_a\}$,
and so we expect the first three gaps of the transfer matrix to be
associated with these operators.

As in Ref.~\cite{djlp}, the scaling dimensions of the operators 
have been extracted by adding to Eq.~(\ref{scal_dims}) a non-universal
$1/L^{4}$ term, and so the fits are based  on two consecutive strip widths $L$.
  
We consider first the case of coupling three identical models.

At the non-trivial fixed point $(x_{2}',x_{3}')= (1,1)$ 
the second and third eigenvalues of the transfer matrix are degenerate. 
These eigenvalues are associated with the operators
$\varepsilon_{\rm A_{1}}$, $\varepsilon_{\rm A_{2}}$
of (\ref{antisym}) which generate the two-dimensional irrep of $S_3$,
as discussed in the Introduction. 

It is important to notice that in the transfer matrix algorithm of
\cite{djlp} the three layers were treated symmetrically and only the
operators which are symmetric under permutation of the layer indices
were accessible. In that case, the degenerate eigenvalues associated
with the energy doublet were not present in the spectrum of the transfer
matrix. The fact that we find them here is a strong confirmation of the
antisymmetric nature of the corresponding operators.

In Table~\ref{tab3} we report numerical values of the dimension
$\Delta(\varepsilon_{\rm A}) \equiv
\Delta(\varepsilon_{\rm A_{1}})=\Delta(\varepsilon_{\rm A_{2}})$
in the case of three coupled Potts models with $q=3$.
The final result $\Delta(\varepsilon_{\rm A}) = 0.63(3)$ is in
reasonable agreement with the RG value and also agrees well with
the Monte Carlo result $\Delta(\varepsilon_{\rm A}) = 0.63 \pm 0.04$
given in \cite{djlp}.

\begin{table}
\begin{center}
\begin{tabular}{r|l}
 $L$     &$\Delta(\varepsilon_{\rm A})$ \\ \hline
 4,6     & 0.6768 \\
 6,8   & 0.6648  \\
 8,10  & 0.6525 \\
 $L\to \infty$ &  0.63(3)\\
 RG result &0.72 \\
\end{tabular}
\end{center}
\caption{\label{tab3}Dimension of the antisymmetric energy
operator at $(x_{2}',x_{3}')= (1,1)$ for three coupled three-state
Potts models.}
\end{table} 
 
The fourth eigenvalue of the transfer matrix was already found in \cite{djlp}
and it corresponds to the symmetric energy operator $\varepsilon_{\rm S}$
of Eq.~(\ref{symm}).
It is a non-trivial check of our transfer matrices that the
corresponding eigenvalue is identical to that of \cite{djlp}.
 
We now turn our attention to the case of three different coupled models.

\begin{figure}
\begin{center}
 \leavevmode
 \epsfysize=400pt{\epsffile{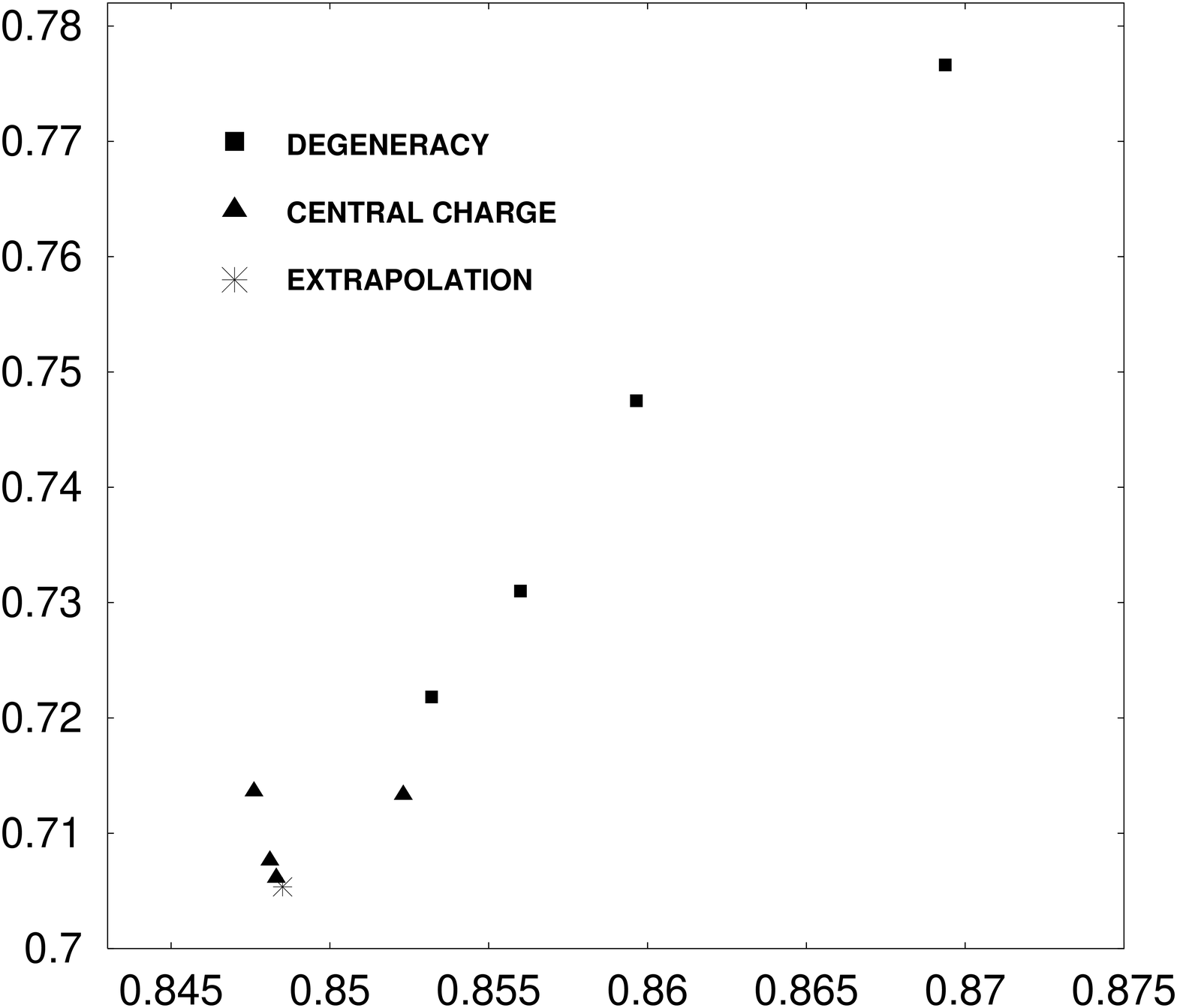}}
 \end{center}
 \protect\caption[3]{\label{fig2}The location of the critical point for the
model ${\cal M}_{345}$, as determined from the maximum of the effective
central charge. Filled triangles show the positions of the maxima
$(x_2'^*,x_3'^*)$ as found from three-point fits with
system sizes ranging from $L=2,4,6$ to $L=8,10,12$. The asterix indicates
the extrapolation of these points to the thermodynamic limit, $L\to\infty$.
Filled boxes give the positions $(x_2',x_3')_{\rm deg}$
of the degeneracy between the second and third
eigenvalues of the transfer matrix, with system sizes ranging from $L=4$
to $L=10$. This latter sequence appears to converge to the same limiting
point as the central charge data, in support of the symmetry restoration
scenario.}
\end{figure}

\begin{figure}
 \begin{center}
 \leavevmode
 \epsfysize=400pt{\epsffile{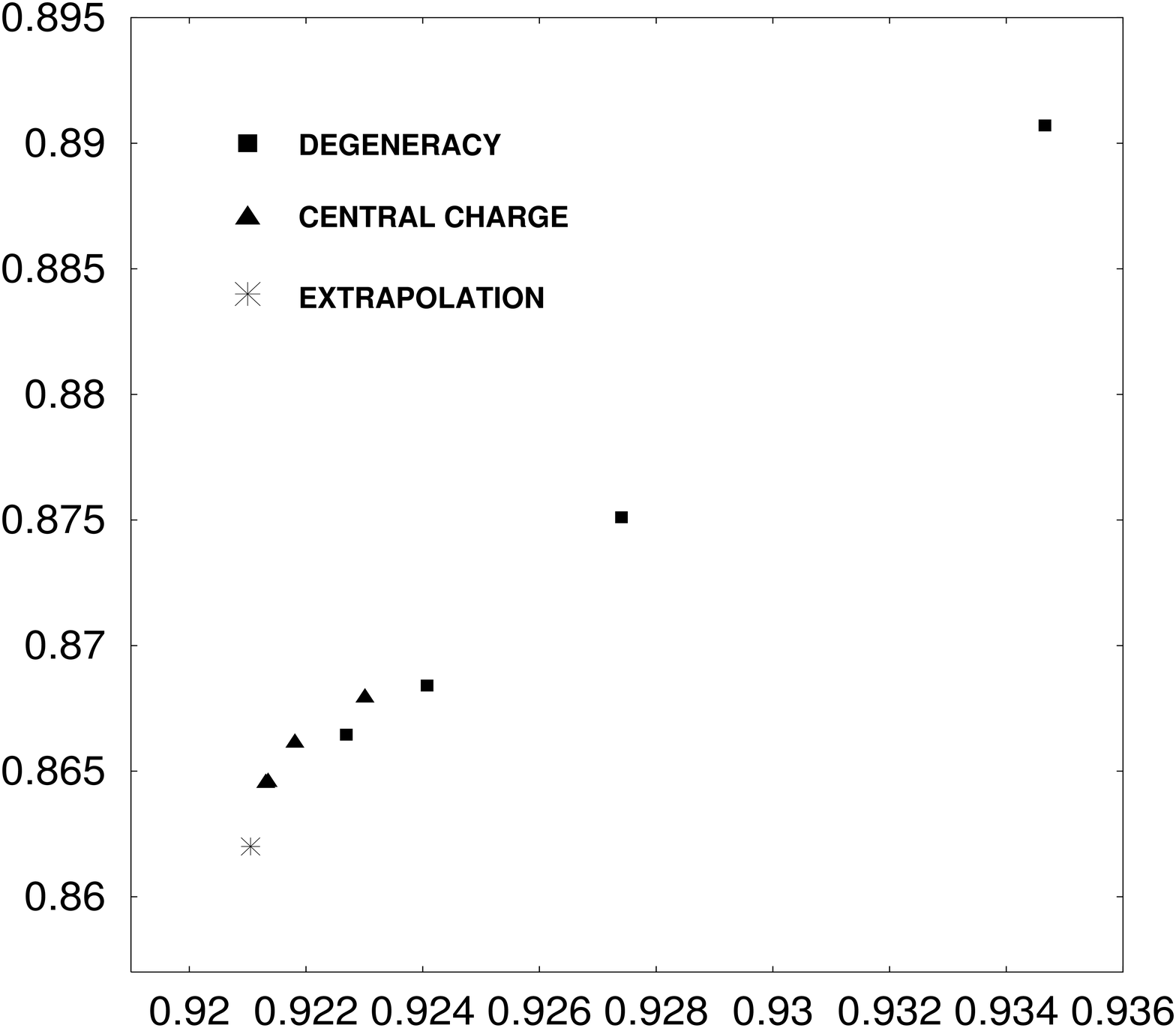}}
 \end{center}
 \protect\caption[3]{\label{fig3}Same data as in Fig.~\ref{fig2}, but here
for the model ${\cal M}_{456}$.}
\end{figure}

In the space $(x_{2}',x_{3}')$ of the couplings we have found 
that for any given system size $L$ there is a unique point
$(x_{2}', x_{3}')_{\rm deg}$, at which the second and third
eigenvalues are degenerate.
We have determined the location of
this point for the models ${\cal M}_{345}$ and ${\cal M}_{456}$ with 
$L=4,6,8,10$ and it is shown on Figs.~\ref{fig2}--\ref{fig3}.

In contradistinction to the case of three coupled identical models, 
the points $(x_{2}'^*,x_{3}'^*)$ and
$(x_{2}', x_{3}')_{\rm deg}$ do not coincide for finite systems.
It is however evident from Figs.~\ref{fig2}--\ref{fig3} that
these two points become closer as the system size increases.
This suggests that in the thermodynamic limit they actually
coincide. If this scenario holds true, we have at the critical
point that $\Delta(\varepsilon_{3}^{*})=\Delta(\varepsilon_{2}^{*})$.

Since the finite-size effects in the sequence of points
$(x_{2}', x_{3}')_{\rm deg}$ are much stronger than is the
case for $(x_{2}'^*,x_{3}'^*)$, it is reasonable to measure numerical
values of the dimensions at the best available estimate for the latter point.
We display the corresponding results for
$\Delta(\varepsilon_{1}^{*})$, $\Delta(\varepsilon_{2}^{*})$,
$\Delta(\varepsilon_{3}^{*})$ in Tables~\ref{tab4}--\ref{tab5}, once
again for the models ${\cal M}_{345}$ and ${\cal M}_{456}$ respectively.
The numerical results are in quite good agreement with the two-loop RG
calculation of section~\ref{sec:RG}. Notice that in the case of
$\Delta(\varepsilon_{2}^{*})$, $\Delta(\varepsilon_{3}^{*})$
we compare with the RG values with the splitting ignored.

\begin{table}
\begin{center}
\begin{tabular}{r|lll}
 $L$     & $\Delta(\varepsilon_{2}^{*})$  & $\Delta(\varepsilon_{3}^{*})$ & $\Delta(\varepsilon_{1}^{*})$\\ \hline
 4,6     & 0.75734  &  0.7677& 1.2991\\
 6,8   & 0.759 &   0.763& 1.2803  \\
 8,10  & 0.7564 & 0.75647 & 1.2761 \\
 RG result & 0.746 & 0.771 & 1.287 \\
\end{tabular}
\end{center}
\caption{\label{tab4}Dimensions of the energy operators at
$(x_{2}'^*,x_{3}'^*)$ for the ${\cal M}_{345}$ model.}
\end{table}
 \begin{table}

\begin{center}
\begin{tabular}{r|lll}
 $L$     & $\Delta(\varepsilon_{2}^{*})$  & $\Delta(\varepsilon_{3}^{*})$ & $\Delta(\varepsilon_{1}^{*})$\\ \hline
 4,6     &0.6849 &0.6857  &1.2958 \\
 6,8   &0.6731  &0.6756   & 1.2765  \\
 8,10  &0.6621  & 0.6637 & 1.2745 \\
 RG result &0.701  &0.728  & 1.43 \\
\end{tabular}
\end{center}
\caption{\label{tab5}Dimensions of the energy operators at
$(x_{2}'^*,x_{3}'^*)$ for the ${\cal M}_{456}$ model.}
\end{table}

In Tables~\ref{tab4}--\ref{tab5} we have given the results for
$\Delta(\varepsilon_{2}^{*})$ and $\Delta(\varepsilon_{3}^{*})$
separately, although we do in fact believe that the two dimensions
coincide. This point of view is corroborated by the fact that
the difference between the dimensions of $\varepsilon_{2}^{*}$
and $\varepsilon_{3}^{*}$ decreases rapidly with the lattice size.
For $L=8,10$ the numerical results give
$\Delta_{\rm splitting}\equiv
|\Delta(\varepsilon_{3}^{*})-\Delta(\varepsilon_{2}^{*})|<0.0003$
for the ${\cal M}_{345}$ model and
$\Delta_{\rm splitting}<0.002$ for the
${\cal M}_{456}$ model. 
% TO DO: Improve for M_456

In order to be sure that the very small values for $\Delta_{\rm splitting}$
obtained at finite size are not somehow accidental, we have determined an
upper bound for $\Delta_{\rm splitting}$ by directly extrapolating
the difference between
the second and third eigenvalues of the transfer matrix at the
point $(x_{2}'^*,x_{3}'^*)$ . The result
is shown below together  with the extrapolated values for the
dimensions of the three energy operators:
\beq
 \begin{array}{llll}
 \mbox{Model ${\cal M}_{345}$}: &
 \ \ \ \ \ \ \ \ \; 
 \Delta(\varepsilon_{1}^{*}) = 1.269 \pm 0.002 & \\ &
 \Delta(\varepsilon_{2}^{*}),\Delta(\varepsilon_{3}^{*}) = 0.750 \pm 0.005
 & \ &
 \Delta_{\rm splitting} < 0.003 \\
 \mbox{Model ${\cal M}_{456}$}: &
 \ \ \ \ \ \ \ \ \; 
 \Delta(\varepsilon_{1}^{*}) = 1.272 \pm 0.002 & \\ &
 \Delta(\varepsilon_{2}^{*}),\Delta(\varepsilon_{3}^{*}) = 0.645 \pm 0.003
 & \ &
 \Delta_{\rm splitting} < 0.002 \\
 \end{array}
\eeq 
The difference $\Delta_{\rm splitting}$ is well below the one predicted
by the RG calculation, which reads respectively $0.025$ and $0.027$ for the
models ${\cal M}_{345}$ and ${\cal M}_{456}$.
The numerical work thus provides clear
evidence that at the non-trivial fixed point the splitting of the
dimensions of $\varepsilon_{3}^{*}$
and $\varepsilon_{2}^{*}$ is actually zero.

\section{Discussion}
\label{sec:disc}

In this paper we have shown that coupling $M=3$ different Potts models
(with $q_1,q_2,q_3>2$ and not too large) one obtains a unique non-trivial
critical point. The critical properties of this point, in particular its
central charge and the values of various energetic scaling dimensions,
have been determined quite accurately by a perturbative RG treatment and
found to be consistent with large-scale numerical simulations.
An exception is however the RG prediction that at two-loop order the
degeneracy between the two antisymmetric energy operators should be
lifted: this prediction has here been discarded on the basis of numerical
evidence.

An extension of our investigation to the case of spin-like operators
will be published elsewhere \cite{djns}.

This work forms part of a larger project \cite{djlp}, in which we examine
the possible universality classes of coupled Potts models, and eventually
their relation to the random-bond Potts model. In particular, substantial
evidence has been accumulated that in the random-bond case replica symmetry
is not broken, and one can thus hope to make analytical progress by studying
the {\em unitary} models that result from coupling a certain number of minimal
models.

We believe that the symmetry properties of the coupled models play
an essential role. It is in the light of this belief that the present work
appears to be interesting: even in the absense of an explicit $S_M$ symmetry
in the initial action, this symmetry appears to be restored at the
non-trivial critical point. As far as a putative CFT classification of
$S_M$ symmetrical critical points goes, we therefore see that the number
of models to be classified is potentially very large. Indeed, assuming
the conclusions of the $M=3$ case to carry over to a general number $M$
of coupled models, one may expect a distinct $S_M$ symmetric universality
class to arise from coupling any different set of $M$ minimal models.
If this is
true, it would call for a substantial number of new CFTs endowed with
extended symmetries. Further research along these lines is currently
in progress.

\end{document}